\documentclass{emulateapj}

\newcommand{\hmpc}{h^{-1}\mathrm{Mpc}}
\newcommand{\hkpc}{h^{-1}\mathrm{kpc}}
\newcommand{\hMsun}{h^{-1}M_{\odot}}
\newcommand{\Msun}{M_{\odot}}

\newcommand{\Omegam}{\Omega_{m}}
\newcommand{\Omegab}{\Omega_{b}}
\newcommand{\Omegal}{\Omega_{\Lambda}}
\newcommand{\band}[2]{\ensuremath{^{#1}\!{#2}}}
\newcommand{\Mr}{M_{\band{0.1}{r}}}
\newcommand{\gr}{\band{0.1}{(g-r)}}
\newcommand{\Mgtot}{M_{g,\mathrm{tot}}}
\newcommand{\Mrtot}{M_{r,\mathrm{tot}}}
\newcommand{\Lstar}{L_\ast}
\newcommand{\Mstar}{M_\ast}
\newcommand{\grtot}{(g-r)_\mathrm{tot}}
\newcommand{\gravg}{(g-r)_\mathrm{avg}}
\newcommand{\grcen}{(g-r)_\mathrm{cen}}
\newcommand{\sigv}{\sigma_v}
\newcommand{\ncen}{n_\mathrm{cen}}
\newcommand{\Mrcen}{M_{r,\mathrm{cen}}}
\newcommand{\Mrsecond}{M_{r,2}}
\newcommand{\Mrgap}{M_{r,\mathrm{2-1}}}
\newcommand{\Rvir}{R_\mathrm{vir}}
\newcommand{\blow}{b_\mathrm{low}}
\newcommand{\bhigh}{b_\mathrm{high}}

\newcommand{\logM}{\mathrm{log}(M/\hMsun)}

\bibpunct[,]{(}{)}{;}{a}{}{,}
\begin{document}

\title{The Clustering of Galaxy Groups: Dependence on Mass and Other Properties}

\author{
Andreas~A.~Berlind,
Eyal~Kazin,
Michael~R.~Blanton,
Sebastian~Pueblas,
Roman~Scoccimarro,
David~W.~Hogg
}
\affil{Center for Cosmology and Particle Physics, New York 
University, New York, NY 10003, USA; aberlind@cosmo.nyu.edu}

\begin{abstract}
We investigate the clustering of galaxy groups and clusters in the SDSS using the 
\citet{berlind_etal_06} group sample, which is designed to identify galaxy systems 
that each occupy a single dark matter halo.  We estimate group masses from their
abundances, and measure their relative large-scale bias as a function of mass.  Our
measurements are in agreement with the theoretical halo bias function, given a
standard $\Lambda$CDM cosmological model, and they tend to favor a low value of the 
power spectrum amplitude $\sigma_8$.  We search for a residual dependence of
clustering on other group properties at fixed mass, and find the strongest signal
for central galaxy color in high mass groups.  Massive groups with less red
central galaxies are more biased on large scales than similar mass groups with redder 
central galaxies.  We show that this effect is unlikely to be caused by errors
in our mass estimates, and is most likely observational evidence of recent 
theoretical findings that halo bias depends on a ``second parameter'' other than mass, 
such as age or concentration.  To compare with the data, we study the bias of 
massive halos in N-body simulations and quantify the strength of the relation
between halo bias and concentration at fixed mass.  In addition to confirming a
non-trivial prediction of the $\Lambda$CDM cosmological model, these results
have important implications for the role that environment plays in shaping galaxy
properties.
\end{abstract}



\section{Introduction} \label{sec:intro}

Our current understanding of galaxy formation is centered on the well-motivated
assumption that all galaxies are formed and live their lives within dark matter
halos.\footnote{Throughout this paper, we use the term ``halo'' to refer to a 
gravitationally bound structure with overdensity $\rho/\bar{\rho}\sim200$, so an 
occupied halo may host a single luminous galaxy, a group of galaxies, or a cluster.  
Higher overdensity concentrations around individual galaxies within a group or cluster 
constitute, in this terminology, halo substructure, or ``sub-halos''.}
The observed clustering of galaxies is thus tied to the clustering of halos.
Indeed, the modern way of modeling galaxy clustering is to combine halo profiles,
abundances, and clustering (measured from cosmological N-body simulations) with 
prescriptions (usually referred to as the halo occupation distribution) that specify 
how galaxies occupy halos (e.g., \citealt{seljak_00,peacock_smith_00,scoccimarro_etal_01,berlind_weinberg_02,cooray_sheth_02}).  
These halo models usually assume that the large-scale clustering of halos depends 
only on halo mass.  Under this assumption, galaxy properties only correlate with 
large-scale environment via their correlation with halo mass.

The dependence of halo clustering on mass has been studied extensively in the context
of $\Lambda$CDM cosmological models and is fairly well understood (e.g., 
\citealt{cole_kaiser_89,mo_white_96,sheth_tormen_99,seljak_warren_04,tinker_etal_05}).
Moreover, the exact form of this dependence is sensitive to the values of cosmological 
parameters, such as the mass density $\Omegam$ and the amplitude of the matter power 
spectrum $\sigma_8$, and can thus, in principle, be used to constrain cosmological 
models.

Recent studies using N-body simulations have shown, however, that the clustering of 
halos also depends on a second parameter, in addition to halo mass.  
\citet{sheth_tormen_04} first noted this when they detected that halos in dense 
environments formed at slightly earlier times than halos of the same mass that are in 
low density environments.  \citet{gao_etal_05} and then \citet{harker_etal_06} 
explored this in a larger simulation and found a much stronger 
signal: early forming halos cluster more strongly on large scales than late-forming 
halos of the same mass.  This effect, sometimes referred to as ``assembly bias'',
was strongest for halos of mass much less than the nonlinear mass $\Mstar$, and it
disappeared for larger masses.  \citet{wechsler_etal_06} did a more thorough analysis
of these issues and, in addition to confirming the \citet{gao_etal_05} result at low 
masses, found the opposite effect for high mass halos, with a turnaround at 
$M\sim\Mstar$.   High mass halos that have recently assembled cluster more strongly 
than old halos of the same mass.  \citet{wechsler_etal_06} also showed that all
these effects persist when using halo concentration, rather than age, as the ``second
parameter''.  This is not surprising given that halo concentration correlates with
formation time \citep{wechsler_etal_02}.  Finally, \citet{wetzel_etal_06} studied
the high mass regime in detail and confirmed the \citet{wechsler_etal_06} results.

The correlation of large-scale environment with halo age/concentration at fixed mass
will propagate into a correlation with galaxy properties if these properties
themselves correlate with halo age/concentration at fixed mass.  \citet{croton_etal_06} 
studied this issue in a semi-analytic model and found that the clustering of red and
blue model galaxies on large scales cannot be explained solely by the color-mass and 
mass-clustering relations.  In other words, galaxy color must also correlate with 
large-scale environment through a ``second parameter'' like halo age or concentration.  
This may not be surprising, but observational analyses have shown that, in the real 
universe, if these effects exist they are quite small
\citep{abbas_sheth_06,skibba_etal_06,blanton_berlind_06}.  The apparent discrepancy 
between the theoretical and observational studies could be due to the fact that
the predicted effects for halos are only strong for either (1) very low mass halos 
that do not contain galaxies luminous enough to be included in the observational
samples, or (2) very high mass halos that are rare and do not contribute much to the
total observed clustering of galaxies.  Alternatively, it is possible that the trends
seen for halos do not, in fact, propagate into trends for galaxies and that the
\citet{croton_etal_06} results do not apply to the real universe.  It is worth noting
that the semi-analytic model used in this study does not predict the correct clustering 
of red and blue galaxies \citep{springel_etal_05}.

A more direct way of observationally probing the theoretical results is to study
not galaxies, but groups and clusters of galaxies, because these are the objects
that presumably correspond one-to-one with halos.  Given a galaxy luminosity limit
near $\Lstar$, which is typical for current large surveys, low mass halos 
($M\sim 10^{12}\Msun$) will contain one isolated galaxy, intermediate mass halos 
($M\sim 10^{13}\Msun$) will contain a group of galaxies, and high mass halos 
($M\sim 10^{14}\Msun$) will contain a cluster.\footnote{Throughout this paper,
we use the term ``groups'' to describe the observational counterparts of halos in all 
three of these regimes.} Studying the clustering of groups can thus serve as a proxy
for studying halos.  In particular, we want to look for an observed group property 
that acts as a ``second parameter'' and shows a clustering dependence at fixed mass.
Many studies have focused on measuring the clustering of groups as a function of 
properties that act as a proxy for mass \citep{bahcall_soneira_83,giurcin_etal_00,bahcall_etal_03b,padilla_etal_04,yang_etal_05b,coil_etal_06}, 
but few have looked at other parameters.  One notable exception is \citet{yang_etal_06},
who found evidence of such an effect using a group catalog derived from the Two 
Degree Field Galaxy Redshift Survey (2dFGRS; \citealt{colless_etal_01}).  Specifically, 
they found that, at fixed estimated group mass, groups containing central galaxies 
with low star formation rates (SFR) are more clustered on large scales than groups 
containing central galaxies with high SFR.  \citet{yang_etal_06} found this result 
for all group masses in their sample.

In this paper we investigate the clustering of groups in the Sloan Digital Sky Survey 
(SDSS, \citealt{york_etal_00}).  We use the \citet{berlind_etal_06} group catalog
that was constructed using an algorithm tuned to identify galaxy systems that each
occupy a single DM halo.  We describe the galaxy and groups samples that we use, as
well as our analysis methods in \S~\ref{sec:data}.  We first study group clustering 
as a function of estimated mass in \S~\ref{sec:groupsmass}, and then search for
a residual dependence on other group properties in \S~\ref{sec:groupsother}.
In \S~\ref{sec:halos} we analyze the clustering of halos in N-body simulations to
compare with our group measurements. We discuss and summarize our results in 
\S~\ref{sec:discussion} and~\ref{sec:summary}.


\section{Data and Analysis} \label{sec:data}

\subsection{SDSS} \label{sec:sdss}

The SDSS is a large imaging and spectroscopic survey that is mapping two-fifths of the 
Northern Galactic sky and a smaller area of the Southern Galactic sky, using a 
dedicated 2.5 meter telescope \citep{gunn_etal_06} at Apache Point, New Mexico.  
The survey uses a photometric camera \citep{gunn_etal_98} to scan the sky 
simultaneously in five photometric bandpasses \citep{fukugita_etal_96,smith_etal_02} 
down to a limiting $r$-band magnitude of $\sim22.5$.  The imaging data are processed 
by automatic software that does astrometry \citep{pier_etal_03}, source identification, 
deblending and photometry \citep{lupton_etal_01,lupton_05}, photometric calibration
\citep{hogg_etal_01,smith_etal_02,tucker_etal_06}, and data quality assessment
\citep{ivezic_etal_04}.  Algorithms are applied to select spectroscopic targets for
the main galaxy sample \citep{strauss_etal_02}, the luminous red galaxy sample
\citep{eisenstein_etal_01}, and the quasar sample \citep{richards_etal_02}.
The main galaxy sample is approximately complete down to an apparent $r$-band
Petrosian magnitude limit of $<17.77$.  Targets are assigned to spectroscopic plates
using an adaptive tiling algorithm \citep{blanton_etal_03a}.  Finally,
spectroscopic data reduction pipelines produce galaxy spectra and redshifts.

Our group sample is identified from the large-scale structure sample 
\texttt{sample14} from the NYU Value Added Galaxy Catalog (NYU-VAGC; 
\citealt{blanton_etal_04a}).  Galaxy magnitudes are corrected for Galactic extinction 
\citep{schlegel_etal_98} and absolute magnitudes are k-corrected 
\citep{blanton_etal_03b} and corrected for passive evolution \citep{blanton_etal_03c}
to rest-frame magnitudes at redshift $z=0.1$.  The galaxy sample that we use
was made publicly available (and superseded) with the SDSS Data Release~4 
\citep{adelman_etal_06}.  We restrict our sample to regions of the sky where the 
completeness (ratio of obtained redshifts to spectroscopic targets) is greater 
than $90\%$.  Our final sample covers 3495.1 square degrees on the sky and 
contains 298729 galaxies.

\subsection{Group and Cluster Samples} \label{sec:samples}

We use the group sample described in \citet{berlind_etal_06}.
Groups were identified using a redshift-space friends-of-friends algorithm
\citep{geller_huchra_83}, which was applied to a volume-limited sample of galaxies 
spanning the redshift range from $z=0.015$ to 0.1 and complete down to an
absolute $r$-band magnitude of -19.9 (the $Mr20$ sample in \citealt{berlind_etal_06}).
This sample contains 57138 galaxies.  With the help of mock galaxy catalogs, the 
group-finding algorithm was tuned to identify galaxy systems that each occupy the same 
underlying dark matter halo.  The result was a group sample that has unbiased 
richness and size distributions with respect to the underlying halo population.
We emphasize that the group-finding algorithm used only the galaxy positions
in redshift space and did not use other properties, such as color.
The volume-limited sample, mock catalogs, group-finding algorithm, and the resulting 
group catalog are described in detail in \citet{berlind_etal_06}.

We compute a total $r$-band luminosity for each group by summing the luminosities
of its member galaxies.  These group luminosities are not actually ``total'', since 
they do not include the light of member galaxies below the -19.9 absolute magnitude
threshold.  However, since the groups were identified from a volume-limited
sample of galaxies, these luminosities should roughly preserve the rank order
of true group luminosities.  We then obtain masses for our groups by matching the 
group luminosity function to a theoretical halo mass function, assuming a monotonic 
relation between group luminosity and mass.  Specifically, we use a 
\cite{warren_etal_06} halo mass function with the following values for cosmological 
parameters: $\Omegam=0.3$, $\Omegab=0.04$, $n_s=1.0$, $\sigma_8=0.9$.  We use this
slightly outdated cosmological model because the mock catalogs that were used to test 
the group-finder were based on this model.  We thus obtain group mass estimates
from their abundances, which \citet{berlind_etal_06} showed to be unbiased with 
respect to the abundances of halos.  Our mass estimates ignore the scatter in the
mass-luminosity relation, but this should not affect our results much because
we only use the group masses to create a set of mass-threshold group samples.
However, we note that these are actually group luminosity threshold samples, which are
characterized as having the same abundance as halos of the stated mass thresholds.

We create four group sub-samples with the following mass thresholds: $\logM>14.0$,
13.5, 13.0, and 12.5.  The resulting number of groups in each of these samples
is 327, 1316, 4299, and 12655, respectively.  Approximately 45\% of groups in
the mass range $12.5<\logM<13.0$ are isolated galaxies (i.e., $N=1$ groups).  On
the high mass end, $\sim 90\%$ of groups more massive than $\logM>14.0$ contain
9 or more galaxies.

We define the position of each group on the sky to be the centroid of its member
galaxy positions, and the group redshift to be the mean member galaxy redshift.  Our
results are not sensitive to these choices, since we focus on the clustering
of groups on large scales.

\subsection{Analysis Method} \label{sec:method}

We wish to measure the relative bias of groups as a function of mass and other
group properties.  The bias between two samples $A$ and $B$ is typically defined 
as the large-scale asymptotic value of $\sqrt{\xi_{AA}(r)/\xi_{BB}(r)}$, where
$\xi_{AA}(r)$ and $\xi_{BB}(r)$ are the two-point autocorrelation functions
of the two samples.  However, our group samples are small enough (especially at
high masses) that the autocorrelations are very noisy, since the number of
group-group pairs is small.  We therefore measure cross-correlations of our
group samples with the full volume-limited galaxy sample from which the groups
were identified.  These cross-correlation functions are much higher signal-to-noise
because the number of galaxies is much higher than the number of groups.  The 
bias between group samples $A$ and $B$ is now the large-scale asymptotic value of
$b(r) = \xi_{AG}(r)/\xi_{BG}(r)$, where $\xi_{AG}(r)$ and $\xi_{BG}(r)$ are the 
cross-correlation functions of group samples $A$ and $B$ with the galaxy sample $G$.
This bias will be identical to the autocorrelation bias, as long as the 
cross-correlation coefficients of the two group samples are the same.

We estimate the cross-correlation function using a symmetrized version of the 
\citet{landy_szalay_93} estimator,

\begin{equation}
\xi_{12} = \frac{D_1D_2}{RR}\left(\frac{n_R^2}{n_1n_2}\right) - \frac{D_1R}{RR}\left(\frac{n_R}{n_1}\right) - \frac{D_2R}{RR}\left(\frac{n_R}{n_2}\right) + 1,
\end{equation}

where $D_1D_2$ are data-data pair counts between the two samples (in this case 
groups and galaxies), $D_1R$ and $D_2R$ are data-random pair counts between each 
sample and a catalog of random points, and $RR$ are random-random pair counts
within the random catalog.  $n_1$, $n_2$, and $n_R$ are the number densities of
group, galaxy, and random catalogs.  Since the group and galaxy samples are both 
volume-limited and have the same geometry, it is sufficient to use one random 
catalog for both.  We use a random catalog containing one million points, which 
is large enough that the noise in the estimated correlation function is dominated 
by the data-data term.  

Since our data are in redshift-space, where galaxy and group peculiar velocities
distort their positions along the line-of-sight, we bin all pairs in a grid
of perpendicular and line-of-sight separations: $r_p$ and $\pi$.  We estimate 
these separations according to \citet{fisher_etal_94}.  For two points with
redshift positions $\mathbf{v}_1$ and $\mathbf{v}_2$, we define the vectors 
$\mathbf{s}=\mathbf{v}_1-\mathbf{v}_2$ and 
$\mathbf{l}=\frac{1}{2}(\mathbf{v}_1+\mathbf{v}_2)$. $r_p$ and $\pi$ for these 
points are then $\pi=(\mathbf{s}\cdot\mathbf{l})/|\mathbf{l}|$ and 
$r_p=\sqrt{\mathbf{s}\cdot\mathbf{s}-\pi^2}$.  Once we estimate $\xi_{12}(r_p,\pi)$, 
we integrate along the redshift direction to get the projected correlation function 
\citep{davis_peebles_83},
\begin{equation}
w_{p,12}(r_p) = 2\int_0^\infty d\pi~\xi_{12}(r_p,\pi).
\end{equation}
In practice, we only integrate out to $\pi_\mathrm{max}=40\hmpc$, following 
\citet{zehavi_etal_02}.

Once we have measured the projected group-galaxy cross-correlation functions for 
two group samples $A$ and $B$, we define a bias function
\begin{equation}
b(r_p) = w_{p,AG}(r_p)/w_{p,BG}(r_p).
\end{equation}
We then compute the asymptotic large scale value of this function by averaging it 
from $r_p=5$ to $20\hmpc$.  This is what we define as the relative bias $b_A/b_B$ 
of the two samples.

\begin{figure}[t]
\epsscale{1.1}
\plotone{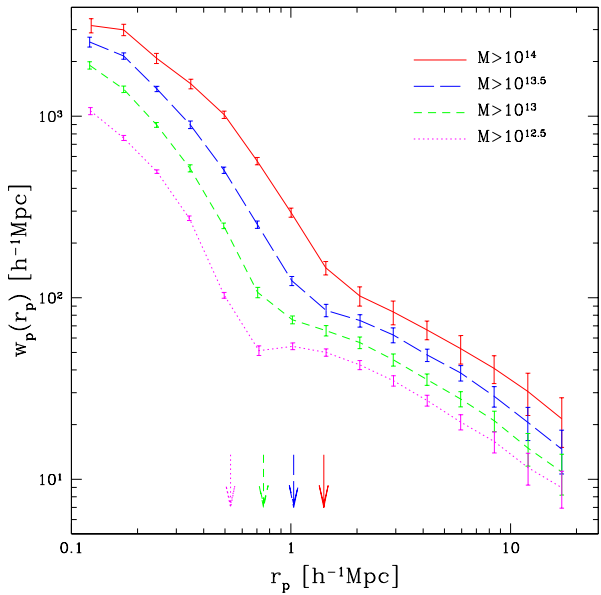}
\caption{
Projected group-galaxy cross-correlation functions for groups of different mass 
thresholds.  Group masses are estimated from their abundances as described in 
\S~\ref{sec:samples}.  The mass thresholds are listed in the panel (in units of 
$\hMsun$) and the vertical arrows denote their virial radii.  The galaxies used
in the cross-correlation have absolute magnitudes $\Mr<-20$.  Errorbars are computed 
by jackknife resampling of the data set.
}
\label{fig:wpall}
\end{figure}

We compute errors by jackknife resampling of the data on the sky 
(e.g., \citealt{zehavi_etal_05}).  We divide our sample area into
twenty, roughly equal-area ($\sim175^\circ$), contiguous regions, and then make our 
measurements twenty times: each time dropping a different region from our samples.  
Since we do not attempt to fit a model to our measurements, it is not necessary to
compute the full covariance matrix.  We only calculate the diagonal errors and
show them in Figures~\ref{fig:wpall} and~\ref{fig:biasMall}.  These errors are 
given by $\sigma_y = \sqrt{\frac{N-1}{N}\sum_{i=1}^{N} (y_i - \bar{y})^2}$, where 
$N=20$, and $y_i$ are a set of $w_p(r_p)$ or $b_A/b_B$ measurements made in the 
twenty jackknife samples.


\section{The Clustering of Groups as a Function of Mass} \label{sec:groupsmass}

Figure~\ref{fig:wpall} shows the group-galaxy projected cross-correlation functions
for our four mass-threshold group samples.  On small scales, this essentially
counts pairs between each group center and its member galaxies.  The 
cross-correlation function on these scales thus simply reflects the average radial 
profile of galaxies within groups.  On scales much larger than the typical size of
groups, the cross-correlation function counts pairs between groups and galaxies 
in other groups.  The transition between these ``1-group'' and ``2-group'' terms
(analogous to the 1- and 2-halo terms in the halo model) is striking in
Figure~\ref{fig:wpall}, much more so than in the galaxy-galaxy autocorrelation
function, where it was detected by \citet{zehavi_etal_04}.

Comparing the different group mass samples on small scales, we see that they
have cross-correlation functions of very similar shape, but more massive 
groups have a higher amplitude and are shifted to larger scales than less massive
groups.  Groups of different masses thus have radial profiles with similar
functional forms, but more massive groups contain a higher overall density of 
galaxies and they extend to larger radii, as expected.  The four vertical arrows
in Figure~\ref{fig:wpall} show the virial radii of our four mass thresholds,
estimated as $\Rvir = (3M/800\pi\bar{\rho})^{1/3}$, where $M$ are the thresholds
and $\bar{\rho}$ is the mean density of the universe (assuming $\Omegam=0.3$).
The 1- to 2-group transitions occur at roughly the virial radii of the mass
thresholds, confirming that (1) the location of the transition reflects the typical
size of groups being considered, and (2) our estimated group masses yield virial
radii that are in agreement with the physical sizes of the groups.

On very small scales ($\lesssim 0.3\hmpc$) the cross-correlation functions flatten,
suggesting that the radial profiles of groups have a core.  However, on these
scales the cross-correlation functions are very sensitive to the definition of
group center.  Any departure of our estimated centers from the ``true'' group
centers would result in just this sort of flattening.

\begin{figure}[t]
\epsscale{1.1}
\plotone{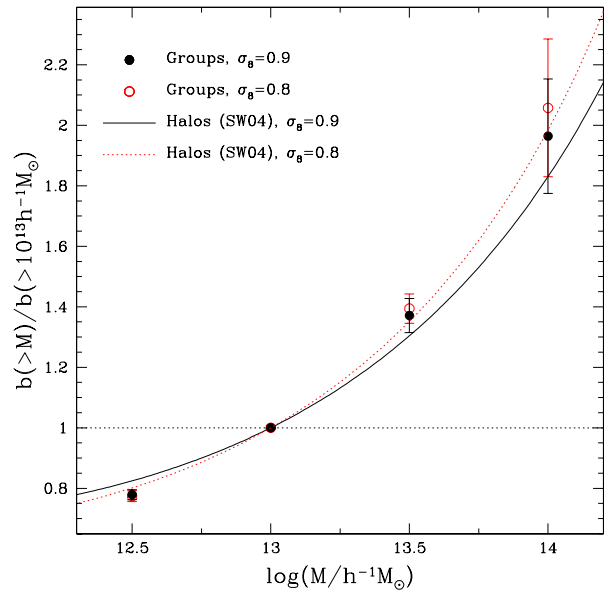}
\caption{
Large-scale bias of groups as a function of group mass threshold, normalized
to the bias of groups with mass greater than $10^{13}\hMsun$.  Data points
are computed by taking the projected galaxy-group cross-correlation function 
of each group mass sample, dividing by that of the $\logM>13$ sample, and 
averaging from 5 to 20$\hmpc$.  Errorbars show the $1-\sigma$ jackknife
uncertainties.  The two sets of points show results for group masses
estimated using two different cosmological models, as listed in the panel.
The two curves show the large-scale bias of dark matter halos as a function
of mass (as computed by \citealt{seljak_warren_04}) for the same cosmological 
models.
}
\label{fig:biasMall}
\end{figure}

On large scales, the relative amplitudes of the cross-correlation functions for 
the different mass groups show that more massive groups are more clustered (i.e.,
have a higher bias) than less massive groups, as expected.  We quantify this
dependence in Figure~\ref{fig:biasMall}, where we show the bias of each sample as 
a function of its mass threshold (filled black points).  We normalize the bias to 
that for the $\logM>13$ sample and thus compute the bias ratio 
$b(M)/b(M=10^{13}\hMsun)$, as described in \S~\ref{sec:method}.  We obtain 
errorbars by jackknife resampling of the data on the sky.  For comparison, 
Figure~\ref{fig:biasMall} also shows the \citet{seljak_warren_04} theoretical bias 
function of dark matter halos, which we also normalize to $M=13\hMsun$ (solid
black curve).  To first order, our measurements agree with the theory.  The
large-scale clustering amplitude of groups rises with mass, following the well-known 
halo bias function 
(\citealt{cole_kaiser_89,mo_white_96,sheth_tormen_99,seljak_warren_04}).

In detail, however, Figure~\ref{fig:biasMall} shows that the bias function for our 
groups rises slightly more steeply with mass than the halo bias function $b(M)$.  
\citet{seljak_warren_04} showed that when halo mass is scaled by the nonlinear
mass $M_*$, the bias function $b(M/M_*)$ is fairly invariant to the values of
cosmological parameters, at least for the limited range of parameter space explored
by those authors (deviations from this were shown by \citealt{tinker_etal_05}).  In 
other words, most of the effect of changing the
cosmological model comes from the corresponding change in $M_*$.  This means that,
in order to have a steeper halo bias function, we must lower $M_*$.  Lowering
$M_*$ will make two fixed masses (e.g., $10^{13}$ and $10^{14}\hMsun$) move to
higher values of $M/M_*$, which will yield a steeper bias ratio between the two
masses.  The cosmological parameter that has the strongest impact on $M_*$
is the power spectrum amplitude $\sigma_8$, with the mass density $\Omegam$ having a
secondary effect and the spectral index $n_s$ placing third.  The red dotted curve 
in Figure~\ref{fig:biasMall} shows the resulting halo bias function when $\sigma_8$ 
is lowered from 0.9 to 0.8.  As expected, the function becomes steeper.  In order to 
be self-consistent, we must now also use this lower value of $\sigma_8$ when we 
assign group masses.  Lowering $\sigma_8$ decreases the abundance of massive halos, 
which leads to lower masses for all groups.  This, in turn, will make our fixed-mass 
threshold samples contain fewer and more rare groups (i.e., higher $M/M_*$) that 
are more strongly clustered.  We calculate new group masses using $\sigma_8=0.8$ 
and re-make our samples.  We show the resulting bias function in 
Figure~\ref{fig:biasMall} (open red points).  The observed group bias function 
becomes slightly steeper, but the change is well within the errorbars and smaller 
than the corresponding change in the theoretical halo bias curve, especially 
for our lowest mass sample.  The theory now gives a better match to the data.

These results demonstrate that, in principle, the combination of group abundances 
and clustering (one to get the mass scale and the other to compare to theory) has 
the power to constrain cosmological parameters (see, e.g., \citealt{mo_etal_96}).  
For example, if we were to take our errors at face value and assume no systematic 
errors, our data would rule out the $(\Omegam,\sigma_8,n_s) = (0.3,0.9,1.0)$ model 
at very high confidence.  Moreover, lowering $\Omegam$ to 0.25 or $n_s$ to 0.95 would 
move in the right direction, but would not be sufficient to give good agreement.  
$\sigma_8$ would have to be lowered as well.  However, this exercise is not very
meaningful without carefully taking into account the main systematic effects present 
in the analysis: (1) our groups do not perfectly correspond to halos, and (2) we are 
ignoring the scatter in the luminosity-mass relation when we assign group masses.  
These issues are especially important given that almost all of the constraining power 
that we have just discussed comes from our lowest mass sample, where most groups only 
contain a couple galaxies.  Properly addressing our systematic errors would require
analysis of realistic mock catalogs constructed using a range of input cosmological 
parameters, in order to understand the biases resulting from group identification, 
and it is beyond the scope of this paper.  The main thing we wish to take away from 
Figure~\ref{fig:biasMall} is that the bias of our groups agrees with that expected 
for halos.  This acts as a good sanity check that our group samples are not too 
different from their underlying halos.  Furthermore, our data seem to prefer a low 
value for $M_*$, in agreement with current CMB constraints \citep{spergel_etal_06}, 
but we cannot say more than this without a detailed treatment of systematic effects.


\section{The Clustering of Groups as a Function of Other Properties} \label{sec:groupsother}

We have established that the large-scale clustering of our groups depends on mass
in the manner expected of dark matter halos.  We now move on to explore the
dependence of clustering on other group properties.  Since many group properties
are likely to correlate with mass, it is possible that any clustering dependence
we find is simply due to this correlation.  Therefore, it is important that we
control for mass and look for residual dependencies on other group properties at
fixed mass.  For each mass threshold sample and for each group property, we split 
the groups into a ``high'' half and a ``low'' half according to that property, in 
a way that keeps the mass distributions of the two halves equal.

The properties we consider are:

(a) $\grtot$ ({\it Total group color}):
We compute a total color for each group by adding up all the $r$-band
light to get $\Mrtot$, adding up all the $g$-band light to get $\Mgtot$, and
setting $\grtot = \Mgtot-\Mrtot$.  This is essentially a luminosity-weighted 
color.  Redder groups make it into the ``high'' samples and bluer groups
make it into the ``low'' samples.

(b) $\grcen$ ({\it Central galaxy color}):
We define the central galaxy to be the most luminous in each group (in the
$r$-band) and use its $\gr$ color.  Groups with redder central galaxies
make it into the ``high'' samples and groups with bluer central galaxies make it 
into the ``low'' samples.

(c) $\gravg$ ({\it Average galaxy color}):
We take the average $\gr$ color of all the galaxies in each group.  This weights
all galaxies equally and thus counts satellite (non-central) galaxies more
than the total group color.  Again, redder groups make it into the ``high'' 
samples and bluer groups make it into the ``low'' samples.

(d) $\Mrcen$ ({\it Central galaxy luminosity}):
This is just the absolute $r$-band magnitude of the most luminous galaxy in each
group.  In cases where $N=1$, $\Mrcen$ is naturally equal to $\Mrtot$.
Groups with a more luminous central galaxy make it into the ``high''
samples and groups with a less luminous central galaxy make it into the ``low'' 
samples.

(e) $N$ ({\it Group multiplicity}):
The number of galaxies with $\Mr<-19.9$ in each group.  We consider all values,
from $N=1$ isolated galaxies, to high $N$ rich clusters.  Groups with more
members make it into the ``high'' samples and groups with fewer members make it 
into the ``low'' samples.

(f) $\ncen$ ({\it Central galaxy concentration}):
A seeing-convolved S\'ersic model is fit to the $i$-band radial light profile
\citep{blanton_etal_03d}.  The S\'ersic index $n$ is a parameter of the fit and
it is a measure of the concentration of the light.  $n=1$ and~4 correspond
to exponential and de Vaucouleurs profiles, respectively.  Values greater than
$n\sim5.9$ are capped at that value.  We use the S\'ersic index of the most
luminous galaxy in each group.  Groups with a more concentrated central galaxy
make it into the ``high'' samples and groups with a less concentrated central 
galaxy make it into the ``low'' samples.

(g) $\sigv$ ({\it Group velocity dispersion}):
This is simply the redshift dispersion of each group, as described in
\citet{berlind_etal_06}.  Groups with $N=1$ have $\sigv=0$.  Groups with a higher 
velocity dispersion make it into the ``high'' samples and groups with a lower
velocity dispersion make it into the ``low'' samples.  

(h) $\Mrgap$ ({\it Luminosity gap}):
This is the magnitude difference between the first and second most luminous
galaxies in each group (in the $r$-band): $\Mrgap = \Mrsecond-\Mrcen$.
In groups with only one galaxy, we use the lower limit for the luminosity
gap, which is the magnitude difference between that galaxy and the magnitude limit 
of our sample: $\Mrgap(N=1) = 19.9 - \Mrcen$.  Groups with a larger gap make it 
into the ``high'' samples and groups with a smaller gap make it into the ``low'' 
samples.  As described in \citet{berlind_etal_06}, galaxies that do not have measured 
redshifts due to fiber collisions were included in the galaxy sample and assigned 
the magnitude and redshift of their nearest neighbor.  As a result, in some groups, 
the first and second most luminous galaxies have the same luminosity.  Since we
have no information about the luminosity gap in these cases, we randomly place half
of them in the ``high'' samples and the other half in the ``low'' samples.

As mentioned above, we wish to study the dependence of large-scale clustering
on these properties {\it at fixed mass} in order to remove any clustering
dependence coming from their correlation with mass.  We do this as follows:
for each candidate group, we define a mass bin of width $\Delta\mathrm{log}M=0.2$
(our results are not sensitive to the choice of bin width)
centered on the mass of the group, and we create a list of all groups whose estimated 
masses lie within the bin.  We then rank this list according to a given group 
property and ask whether our candidate group sits in the top or bottom 50\% of the 
ranked list.  If it is in the top 50\% we place it into the ``high'' sample, and if 
it is in the bottom 50\% we place it into the ``low'' sample.  In this way, we create
two samples of roughly equal size that have identical mass (total luminosity)
distributions, but quite disjoint distributions in the group property used to
split them.  We do this separately for each of the properties in the above list.

\begin{figure}[t]
\epsscale{1.2}
\plotone{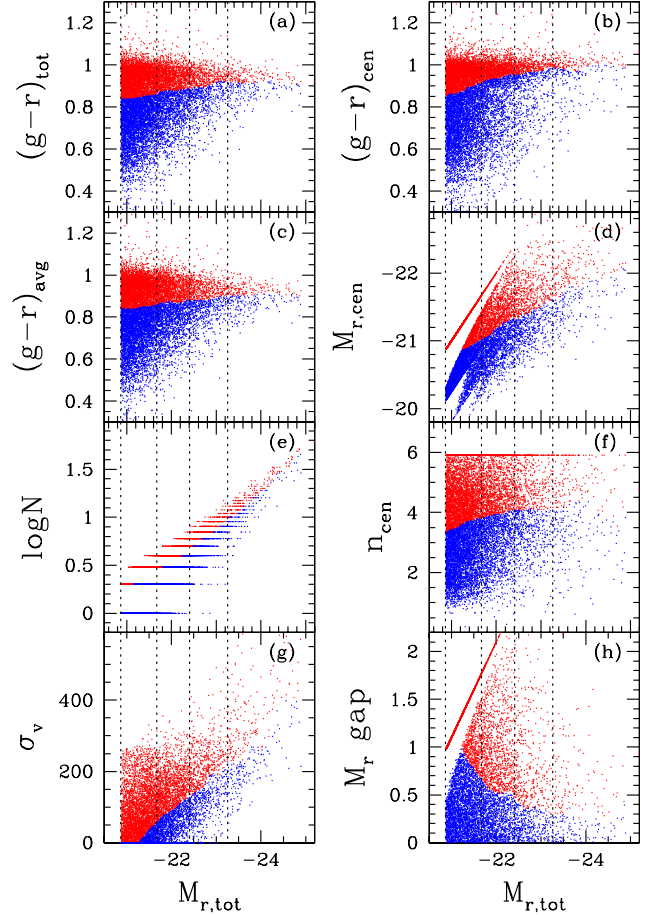}
\caption{
Group properties as a function of total absolute magnitude, showing the
divisions into two samples for each property.  Each panel shows a
particular group property: (a) total group $g-r$ color, (b) brightest
galaxy $g-r$ color, (c) mean galaxy $g-r$ color, (d) brightest galaxy
$r$-band absolute magnitude, (e) group multiplicity, (f) brightest
galaxy sersic index, (g) group velocity dispersion, (h) absolute
magnitude gap between first and second brightest galaxy.  For each
property, groups are placed into a ``top 50\%'' (red dots) or 
``bottom 50\%'' (blue dots) sample depending on whether their value 
is higher or lower than the median value for all groups at fixed group mass.
Vertical dotted lines show the total absolute magnitudes corresponding
to our four group mass thresholds.
}
\label{fig:properties}
\end{figure}

Figure~\ref{fig:properties} shows the result of this splitting procedure.  Each
panel shows the relation between a group property and $\Mrtot$ (our proxy for mass).
Groups that were placed in the ``high'' and ``low'' samples are represented by
red and blue dots, respectively.  The four vertical dotted lines show the group
luminosities that correspond to our four mass thresholds.  The boundary between
the two samples in each case shows the median relation of each group property with
total group luminosity.  Some properties, like central galaxy luminosity, group
multiplicity, and velocity dispersion, have strong correlations with group luminosity
for all $\Mrtot$, whereas others, such as the three colors, central concentration, 
and luminosity gap, correlate with group luminosity only at low $\Mrtot$.  The
three properties that correlate the most with $\Mrtot$ are simple to understand.
$\Mrcen$ correlates with $\Mrtot$ because the total group luminosity includes the 
central galaxy luminosity and is often dominated by it in small groups.  At fixed
$\Mrtot$, however, we have no reason to think that $\Mrcen$ will correlate with
mass.  In other words, $\Mrtot$ and $\Mrcen$ probably do not correlate with mass
independently of each other.  The same is not true for group multiplicity or 
velocity dispersion.  These properties likely correlate with group mass more
independently of $\Mrtot$.  For example, at fixed group luminosity, groups with
higher velocity dispersion probably have larger masses on average than groups
with lower velocity dispersion.  Another way to say this is that $\sigv$ and $N$
probably correlate with the scatter in mass at fixed $\Mrtot$.

\begin{figure}[t]
\epsscale{1.1}
\plotone{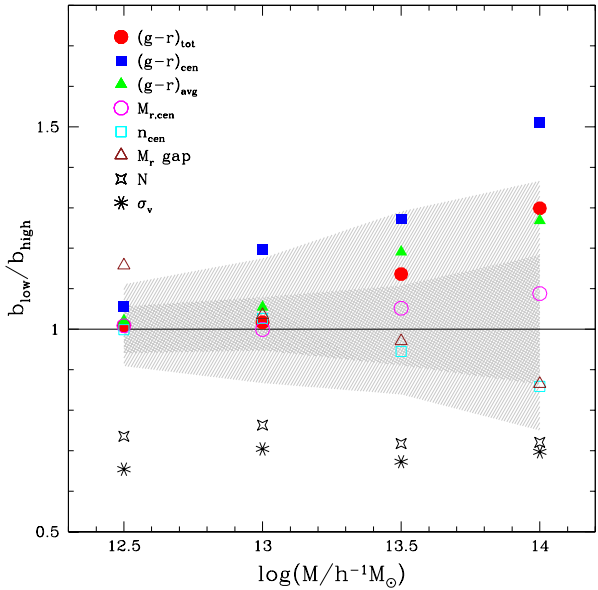}
\caption{
Ratio of large-scale bias of groups with low values of a given property over
that of groups with high values of that property, as a function of group mass
threshold.  At each mass threshold, different points show this bias ratio for 
groups split by different properties.  Point types and properties are listed 
in the panel.  All points are computed by taking the projected galaxy-group 
cross-correlation function of the ``low'' sample, dividing by that of the 
``high'' sample, and averaging from 5 to 20$\hmpc$.  The shaded regions enclose 
68\% (dark shaded region) and 95\% (light shaded region) of the values of 
$\blow/\bhigh$ that result from randomly splitting the group sample into two 
halves 200 independent times, and thus represent $1-$ and $2-\sigma$ detection levels.
}
\label{fig:biasMredblue}
\end{figure}

Now that we have split the groups into halves according to each group property,
we can measure the bias ratio $\blow/\bhigh$ (as described in \S~\ref{sec:method})
for each property as a function of group mass.  The results are shown in 
Figure~\ref{fig:biasMredblue}.  Each type of point corresponds to a particular
group property, as listed in the panel.  Values of $\blow/\bhigh$ equal to unity
indicate that the large-scale clustering of groups does not depend on that
group property.  Values above unity indicate that groups in the ``low''
sample cluster more strongly than those in the ``high'' sample, and the opposite
is true for values below unity.

In order to assess the significance of our results, we create 200 realizations
of random ``high'' and ``low'' samples, where the group property used in the
splitting is a random number.  The dark and light shaded regions in 
Figure~\ref{fig:biasMredblue} enclose 68\% and 95\% (i.e., $1\sigma$ and $2\sigma$)
of these 200 realizations, respectively and are centered on the median.  The shaded 
band grows with mass because the higher mass samples contain fewer groups and their 
clustering measurements are therefore noisier.  We have also computed jackknife 
errors for our $\blow/\bhigh$ measurements and verified that they are roughly in 
agreement with the random splitting errors.

We draw several conclusions from Figure~\ref{fig:biasMredblue}:

(1) Only group multiplicity and velocity dispersion correlate significantly with
large-scale clustering at all estimated group masses.  Furthermore, they behave in
exactly the way that is expected if these group properties correlate with mass
at fixed group luminosity.  Groups with higher $N$ or $\sigv$ cluster more strongly
than groups with lower $N$ or $\sigv$.  Moreover, the bias ratio $\blow/\bhigh$ 
is approximately the same ($\sim 0.7$) for both properties and for all masses.
If $N$ and $\sigv$ correlate with mass at fixed group luminosity, as we expect,
then their bias ratio will depend on two factors: the amount of scatter in the 
luminosity-mass relation (after all, if the scatter is zero then $N$ and $\sigv$
cannot provide additional information on the mass), and the steepness of the bias 
function $b(M)$ (the steeper the function, the more a given amount of scatter can 
affect the bias).  These two effects have opposite dependencies on mass: as mass 
increases, the scatter in the luminosity-mass relation should decrease, while the 
bias function steepens.  Figure~\ref{fig:biasMredblue} suggests that these two 
effects balance each other out to produce a roughly constant $\blow/\bhigh$ ratio 
as a function of mass.

(2) The three group colors that we consider show evidence of a correlation with
large-scale clustering at high mass, but not at low mass.  Of these, the most 
significant is the central galaxy color.  High-mass groups and clusters with 
blue central galaxies are more strongly biased than those with red central galaxies.  
Actually, at these masses almost all central galaxies are red (see the second
panel in Fig.~\ref{fig:properties}), so it is more accurate to say that groups with
less red central galaxies are more clustered than those with redder central
galaxies.  This is a $\sim2\sigma$ effect for the $10^{13}$ and $10^{13.5}\hMsun$ 
mass thresholds and grows to a $\sim3\sigma$ effect at $10^{14}\hMsun$.
There is a similar effect for total and average group colors in the sense that
less red massive groups are more biased than redder groups, but it is only
at the $\sim1.5\sigma$ level of significance.  Moreover, group color correlates
with central galaxy color by construction, so it is not clear how much of the 
signal seen for total and average group color actually comes from the color of the
central galaxy.

(3) There is a hint that, at high masses, groups with more concentrated central
galaxies are more biased than groups with less concentrated central galaxies.
This may seem to be in conflict with the result for central galaxy color because
usually more concentrated galaxies are redder.  However, at these very high
luminosities, no such correlation exists.  In any case, this is only a $1\sigma$
measurement, so we do not consider it significant.

(4) The luminosity gap statistic shows that groups with smaller gaps are more 
biased at low masses and less biased at high masses.  At high masses this is only 
a $1\sigma$ effect, but at the lowest mass threshold it is a $\sim3\sigma$ effect.  
We must be cautious, however, because at these lowest masses, a large fraction of 
groups have $N=1$, where the gap statistic is highly unreliable.  There could 
therefore be a large systematic error that affects this point.  We test this by 
repeating the measurement after throwing out all $N=1$ groups, and we find that the 
detection disappears completely.  The overall significance of the luminosity gap 
detections are therefore low.

(5) Central galaxy luminosity does not appreciably correlate with large-scale
clustering at fixed total group luminosity.  Since we are comparing samples
at fixed total group luminosity, splitting by central galaxy luminosity is
the same as splitting by the ratio of central-to-total luminosity.  This ratio
is similar to the luminosity gap statistic in the sense that groups with a high 
central-to-total luminosity ratio are likely to also have a large luminosity gap.
However, the central-to-total ratio is more robust than the luminosity gap because
it is less affected by systematic effects due to incompleteness and group 
misidentification.  The different behavior exhibited by these two statistics
provides additional evidence that the luminosity gap results are not significant.

Aside from the observed clustering dependence on group multiplicity and velocity 
dispersion, which are most likely due to their correlation with mass,
our most significant detection is that group color (especially the central
galaxy color) correlates with large-scale bias in high luminosity groups.
The simplest explanation for this effect is that group color correlates with
mass at fixed group luminosity, much like multiplicity and velocity dispersion.
One potential problem with this explanation is that the observed bias ratio for 
central galaxy color is more extreme than that for velocity dispersion, which 
implies that color must correlate more strongly with mass than does $\sigv$.  This 
definitely seems odd, but perhaps it is true.  Since we know that velocity
dispersion should correlate with mass at fixed group luminosity, we can check
whether groups with less red central galaxies have a different $\sigv$ distribution
than groups with redder central galaxies.

\begin{figure}[t]
\epsscale{0.93}
\plotone{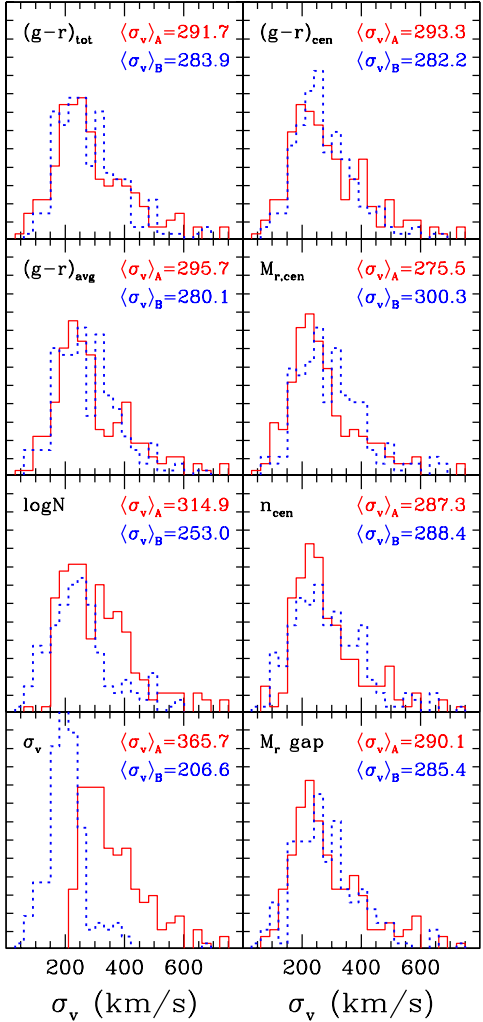}
\caption{
Velocity dispersion histograms for samples of massive ($M>10^{14}\hMsun$) groups split 
by different group properties.  Each panel shows a particular group property,
as in Fig.~\ref{fig:properties}.  In each panel, the solid and dotted histograms 
show the velocity dispersion distribution for groups with high and low values of 
the property, respectively.  The mean velocity dispersion for the ``high'' and
``low'' samples are denoted $\langle\sigma_v\rangle_A$ and $\langle\sigma_v\rangle_B$,
respectively, and are listed in each panel.
}
\label{fig:sigvhist}
\end{figure}

Figure~\ref{fig:sigvhist} shows the velocity dispersion distributions for the
``high'' (solid red histograms) and ``low'' (dotted blue histograms) samples 
for each group property in the case of $\logM>14$ groups.  Each panel also
lists the mean values of $\sigv$ for the two samples.  With the exception of
multiplicity and, of course, velocity dispersion itself, all other properties
have ``high'' and ``low'' samples with nearly identical distributions.
In other words, these properties show no correlation with $\sigv$.  This
result provides strong evidence that whatever trends exist in
Figure~\ref{fig:biasMredblue} are not due to correlations with mass.  The one
notable exception is group multiplicity, which correlates with $\sigv$, as
expected.

We discuss the implications of these results in \S~\ref{sec:discussion}.

\section{The Clustering of Halos} \label{sec:halos}

In the previous section, we showed that there is a significant dependence of 
high mass group clustering on a group property other than mass, and that this
dependence is most likely not due to a correlation with mass.  One possibility
is that this is a detection of the recently discovered assembly bias for dark 
matter halos; namely, that halo clustering depends on halo age/concentration 
at fixed mass.  \citet{wechsler_etal_06} found that for $M>\Mstar$, low
concentration halos are more strongly biased than high concentration halos.
However, the simulations used in that study were fairly small in volume
(albeit very high in resolution) making it difficult to study the high mass
regime where halos are rare.  By necessity, the \citet{wechsler_etal_06} 
results for high $M/\Mstar$ came from high redshift outputs of the simulation, 
where $\Mstar$ is lower.  \citet{wetzel_etal_06} investigated this with a much 
larger simulation and confirmed the \citet{wechsler_etal_06} results for high 
mass halos.

We wish to compare the age/concentration effect for high mass halos to our
results for galaxy groups.  We therefore need to know the ratio $\blow/\bhigh$
for halos, where halos are split into ``low'' and ``high'' samples in a similar 
way as groups were split in \S~\ref{sec:groupsother}.  We do this using a set
of N-body simulations, which we describe below.

\subsection{N-body Simulations and Analysis} \label{sec:nbody}

We use 20 independent N-body simulations of a $\Lambda$CDM cosmological model, 
with $\Omegam=0.27$, $\Omegal=0.73$, $\Omegab=0.046$, 
$h\equiv H_0/(100~\mathrm{km~s}^{-1}~\mathrm{Mpc}^{-1})=0.72$, $n_s=1.0$, and
$\sigma_8=0.9$.  Each simulation follows the evolution of $512^3$ dark
matter particles of mass $m_p=7.49\times10^{10}\hMsun$ in a periodic box of length 
$512\hmpc$.  Initial conditions were set up using second-order Lagrangian 
perturbation theory (2LPT) as described by \citet{crocce_etal_06}, and the simulations 
were run using the GADGET2 code \citep{springel_05}.  The gravitational force 
softening was set to $\epsilon_{\rm grav}=40\hkpc$ (Plummer equivalent).  The 
simulations are described in more detail in \citet{crocce_etal_06}.

We identify halos in the $z=0$ dark matter particle distributions using a 
friends-of-friends algorithm with a linking length equal to $0.2$ times the
mean inter-particle separation.  The mass of each halo is set to the sum
of its member particle masses, times a correction factor given by 
\citet{warren_etal_06} that accounts for systematic effects due to the finite
number of particles per halo.  For the large halo masses that we consider in 
this paper, this correction is never larger than 1.5\%.  This procedure yields 
a total of 90231 halos (in all 20 simulations) of mass greater than $10^{14}\hMsun$.  
For each halo, we compute a virial radius equal to 
$\Rvir=\left(\frac{3M}{800\pi\bar{\rho}}\right)^{1/3}$, where $M$ is the halo mass 
and $\bar{\rho}$ is the mean density of the universe.  Finally, we assign 
center-of-mass coordinates to all halos.

Halo concentrations are simpler to measure than ages because they can be measured 
from the $z=0$ simulation outputs alone and do not require the construction of
merger trees.  For this reason, we use concentrations in this study; however,
we note that similar results should hold for formation time since these halo
properties correlate with each other \citep{wechsler_etal_02}.  We measure a 
concentration for each halo in the simplest possible way: we choose a fixed 
fraction of the halo virial radius and measure the ratio of mass enclosed within 
this radius to the total mass.  We choose a value for this fraction equal to 0.3, 
but we also compute concentrations using values as low as 0.1 and as high as 0.5 
in order to test how robust our results are to this definition.

We then split all halos of mass greater than $10^{14}\hMsun$ into ``high'' and ``low''
samples, in exactly the same way as we split our groups in \S~\ref{sec:groupsother}.
For each halo, we create a mass bin of width $\Delta\mathrm{log}M=0.1$ that is 
centered on that halo's mass, and we make a list of all halos that lie within
this bin.  We then sort this list by concentration, and we place the halo
in question in the ``high'' or ``low'' sample depending on whether its concentration
places it in the top or bottom 50\% of the sorted list.  We also create a second set 
of samples by choosing halos whose concentrations place them in the top or bottom
25\% of halos within their mass bin.  In this way, we create halo samples of high
and low concentration that have the same mass distributions, thus controlling for
the well-known concentration-mass relation \citep{bullock_etal_01}.

We measure halo-halo autocorrelation functions for the ``high'' and ``low'' samples
in each of the 20 simulations.  We use the simple estimator $\xi(r) = DD/RR - 1$,
where $DD$ is the number of halo-halo pairs, and $RR$ is the number of random-random
pairs, which we calculate analytically as
$RR = (1/2)(4/3)\pi\bar{n}^2 V_{\mathrm{box}} (r_{\mathrm{out}}^3-r_{\mathrm{in}}^3)$,
where $\bar{n}$ is the mean density of halos, $V_{\mathrm{box}}$ is the volume
of the simulation cube, and $r_{\mathrm{out}}$ and $r_{\mathrm{in}}$ are the outer
and inner radii of the bin in which $\xi(r)$ is being calculated.  We then
calculate bias functions $b(r) = \sqrt{\xi_\mathrm{low}(r)/\xi_\mathrm{high}(r)}$.
We compute average correlation functions and bias functions from the 20 simulations,
and we calculate errors from the dispersion among these independent realizations.

\subsection{Results} \label{sec:nbodyresults}

\begin{figure}[t]
\epsscale{1.0}
\plotone{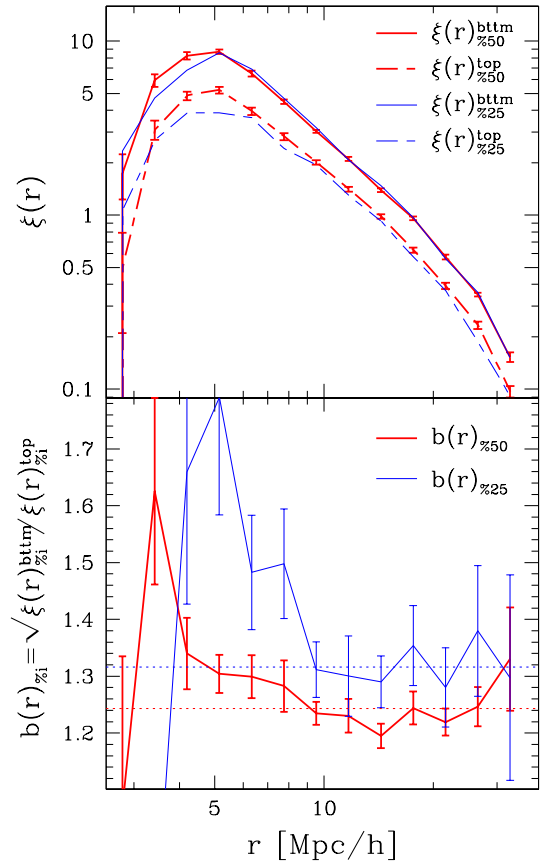}
\caption{
{\it Top panel}:
Two-point autocorrelation functions for simulated dark matter halos of mass 
greater than $10^{14}\hMsun$, split into subsamples by concentration.  Thick 
red solid and dashed curves show correlation functions of the 50\% least and
most concentrated halos, respectively.  Thin blue solid and dashed curves 
show the same for the 25\% least and most concentrated halos.
{\it Bottom panel}:
Bias function defined as the square root of the correlation function for low 
concentration halos divided by that for high concentration halos.  The thick
red, and thin blue curves shows the bias for the 50\% and 25\% concentration 
splits, respectively.  Horizontal doted lines show the values of the bias,
averaged from 9 to 35$\hmpc$ in these two cases.  The values of this large-scale
bias are $1.243\pm 0.017$ and $1.316\pm 0.032$ for the 50\% and 25\% concentration 
splits, respectively.
In both panels, results are averaged over 20 independent N-body simulations,
and errors show the uncertainty in the mean estimated from the dispersion
among these simulations.
}
\label{fig:xibias}
\end{figure}

The top panel of Figure~\ref{fig:xibias} shows halo correlation functions for
the 50\% ``low'' and ``high'' samples (thick red curves), as well as the
25\% ``low'' and ``high'' samples (thin blue curves).  All correlation functions
show the expected turnover at low scales due to the fact that friends-of-friends
halos cannot overlap.  There can thus be no halo pairs at separations less than
twice the virial radius of the smallest halos considered.  The bottom panel shows 
low/high bias functions for the 50\% (thick red curve) and 25\% (thin blue curve) 
samples.  

Figure~\ref{fig:xibias} clearly shows that low concentration halos have a 
significantly higher amplitude of $\xi(r)$ than high concentration halos,
at fixed mass.  This is a very high signal-to-noise confirmation of the
\citet{wechsler_etal_06} and \citet{wetzel_etal_06} results.  The bias functions
become scale-independent at large scales, showing that halo concentration affects
the amplitude, but not the shape of the correlation function, as expected.
The low/high bias of the 25\% samples is naturally higher than that of the 50\% 
samples, but it is interesting that all the difference comes from the high 
concentration end.  The correlation functions of the lowest 50\% and 25\%
concentrated halos have exactly the same amplitude.  This means that the relation
between concentration and halo bias, at fixed mass, is flat below the median
concentration, and only starts dropping at higher concentrations.

We compute large-scale asymptotic values of $\blow/\bhigh$ by averaging the
bias functions in the range where they are scale independent: from 9 to $35\hmpc$.
We do this separately for each of the 20 simulations and then compute the
mean and uncertainty in the mean, which we calculate from the dispersion
among the simulations.  The relative bias of low over high concentration halos 
is $\blow/\bhigh = 1.243\pm0.017$ for the 50\% samples and 
$\blow/\bhigh = 1.316\pm0.032$ for the 25\% samples.  These values are shown in
Figure~\ref{fig:xibias} as dotted lines.  We note that our results are not sensitive 
to the specific definition of concentration that we use.  When we measure concentrations
using different fractions of the virial radius, the values of $\blow/\bhigh$ change
by less than their uncertainties.  In the next section, we discuss the connection 
between these results for halos and those for groups shown in the previous section.

\section{Discussion} \label{sec:discussion}

In \S~\ref{sec:groupsother}, we found that the large-scale bias of massive groups
correlates with their central galaxy color, at fixed estimated group mass.  
Specifically, massive groups with less red central (brightest) galaxies cluster more 
strongly than those with redder central galaxies.  Furthermore, we showed that this 
trend is probably not due to a correlation of central galaxy color with true mass at 
fixed estimated mass.  In other words, we have found that the large-scale clustering
of massive groups depends on a ``second parameter'' other than mass.  This result can 
be understood in the context of the recent findings by \citet{wechsler_etal_06} and 
\citet{wetzel_etal_06}, as well as our results in \S~\ref{sec:halos}, showing that 
massive dark matter halos also have a ``second parameter'', whether it be
concentration, age, or some other feature of the halo assembly history.  
If central galaxy color correlates with halo concentration or age at fixed halo mass,
then it will also correlate with large-scale bias, since halo concentration and age 
correlate with bias.  Moreover, the effect should only be present at high masses,
and should disappear as $M$ approaches $\Mstar$, which is exactly what we 
find.\footnote{The effect should re-appear with the opposite sign at lower masses, but
our group catalog does not probe that mass regime.}  This means that our result for 
groups most likely constitutes observational evidence for the recently discovered 
``second parameter'' effect for halos.  The only alternative is that the color of a 
halo's central galaxy somehow ``knows'' about the density field at large scales 
($\sim10\hmpc$), independently of its halo's assembly history.  Though possible in 
principle, this is highly unlikely.

For groups of mass greater than $10^{14}\hMsun$, we found that the relative bias
between groups with the bottom and top 50\% of central galaxy color is 
$\blow/\bhigh\sim1.5\pm0.2$.\footnote{The uncertainty comes from jackknife resampling.}
For comparison, the relative bias between the 50\% low and high concentration halos 
in our simulations is $\blow/\bhigh=1.243\pm0.017$.  This bias ratio is closer to
what we found for total and average group color.  This difference could indicate 
that central galaxy color correlates with a feature of the halo assembly history that 
is more directly tied to large-scale bias than concentration.  However, the difference 
between these numbers is not highly statistically significant.  This highlights the 
need for repeating these observational investigations with a much larger sample 
volume that contains a much larger number of massive groups.  

Our results imply that massive halos that formed earlier contain redder central 
galaxies than halos of the same mass that assembled more recently.  This is
interesting because it is not obvious what one should expect.  One could argue
that recently assembled halos are more likely to have had a recent major merger, 
which might result in a central galaxy with no star formation, whereas older halos 
might contain central galaxies that have had time since their last merger to accrete
gas and have some ongoing star formation.  On the other hand, one could argue that
in this high mass regime, there will be no cool gas to accrete and that older halos
will simply have older central galaxy stellar populations.  Our results suggest the
latter scenario.  

It is interesting that the luminosity gap, which we expected to be the group property 
most directly related to age, showed no significant correlation with large-scale bias 
at fixed mass.  The reasoning behind our expectation is that in old groups, satellite 
galaxies will have had more time to merge with the central galaxy, resulting in a very 
massive central galaxy and thus a large luminosity gap.  If anything, our results show 
the opposite trend than expected: massive groups with a high luminosity gap (and hence
older) are slightly more clustered than groups with a low gap.  This contradicts the
theoretical results for halos.  One possible explanation for this is that the 
luminosity gap does not actually correlate with age in the way we expect.  However, it 
could also be that our measured luminosity gaps suffer severely from systematic effects 
due to incompleteness and group misidentification and are, therefore, not representative 
of the true luminosity gaps within {\it halos}.  We believe this latter explanation 
because we find quite different results when we study the dependence of clustering on 
central galaxy luminosity, which should act like the luminosity gap.  Old groups
should have both a larger luminosity gap, and a more luminous central galaxy than young 
groups of the same mass.

\citet{zentner_etal_05} predicted that halo concentrations and formation times should
be correlated with the number of subhalos within the halo in the sense that older,
more concentrated halos have fewer subhalos.  This correlation arises because older
halos accrete their subhalos earlier and thus give them more time to sink to the
center and be destroyed.  If this prediction is correct, we should see a trend whereby 
high mass groups with many member galaxies (high $N$) are more strongly biased than 
poorer groups of the same mass.  Moreover, this trend should disappear at smaller group 
masses since the bias-age trend also goes away at smaller masses.  In this paper, we 
find the predicted trend at high masses, but, unlike the prediction, it persists at 
lower masses.  It is not straightforward to interpret our result in terms of the 
\citet{zentner_etal_05} prediction because we do not compare groups at fixed mass,
but rather at fixed total luminosity.  If group multiplicity $N$ correlates with
group mass at fixed group luminosity then we also expect a trend like the one we find.
Since the trend we see persists at low group masses, we conclude that we are likely
detecting a multiplicity-mass correlation rather than an intrinsic multiplicity-bias
relation. 

\citet{yang_etal_06} have also detected a ``second parameter'' effect for groups.
In a study similar to ours, but using 2dFGRS groups, they found that groups containing
central galaxies with low SFR are more biased than groups containing central galaxies
with high SFR.  Moreover, they found this effect at all group masses.  This seems
to be at odds with our results because if low SFR implies redder color, then we
have found the opposite effect.  \citet{yang_etal_06} use the \citet{madgwick_etal_02}
$\eta$ parameter as a proxy for SFR, which reflects the average emission- and 
absorption-line strength in the galaxy rest frame spectrum.  This is not necessarily
correlated with $g-r$ color for the luminous red galaxies that are the central
galaxies of these massive systems.  Nevertheless, at its face, this result looks
completely opposite to ours and remains an interesting puzzle.


\section{Summary} \label{sec:summary}

In this paper we have investigated the clustering of galaxy groups in the SDSS.
The principal goal of this study was to look for secondary dependences of large-scale 
clustering on group properties other than mass.  We have estimated group masses
from their abundances using the total group luminosity as a proxy for mass.  Our main 
results are:

1.
The measured large-scale bias of groups as a function of estimated mass is in
agreement with the theoretical halo bias function, given a standard $\Lambda$CDM
cosmological model.  The measurements suggest a preference for a low value of 
$\sigma_8$, in agreement with current CMB constraints, but this result is not
robust since it could be subject to systematic effects.

2.
We have measured the residual dependence of group bias on other group properties,
at fixed estimated mass.  The properties we have considered are: total group color,
central galaxy color, average group color, central galaxy luminosity, group
multiplicity (richness), central galaxy concentration, group velocity dispersion,
luminosity gap between first and second brightest galaxies.  Of these properties,
only group multiplicity, velocity dispersion, and central galaxy color show a
significant correlation with bias at fixed estimated mass.  Specifically, groups
with higher multiplicity, higher velocity dispersion, and less red central galaxies
cluster more strongly than groups with the opposite properties.  The effect for
multiplicity and velocity dispersion occurs at all masses, whereas the effect
for central galaxy color is only significant at high group masses ($M>10^{14}\hMsun$).

3.
The dependence of large-scale bias on group multiplicity and velocity dispersion can 
be simply explained if these properties correlate with true mass at fixed estimated 
mass.  However, the dependence of bias on central galaxy color cannot be explained 
this way and is most likely a real effect.  This is likely observational evidence of
recent theoretical findings that halo bias depends on a ``second parameter'' other than
mass, such as age or concentration.

4.
Our results imply a connection between halo age and central galaxy color for massive 
halos.  Halos that assembled earlier likely contain redder central galaxies than
recently assembled halos of the same mass.

5.
In order to compare our results to theory, we have quantified the dependence of halo 
bias on concentration for high mass halos, using a set of large N-body simulations.  
We find that low concentration halos are more biased on large scales than high 
concentration halos, at fixed mass, thus confirming previous results at very high
signal-to-noise.  Specifically, we find that for halos with $M>10^{14}\hMsun$, the 
relative bias between the 50\% least and most concentrated halos is 
$\blow/\bhigh = 1.243\pm0.017$, while the relative bias between the 25\% least and
most concentrated halos is $\blow/\bhigh = 1.316\pm0.032$.


\acknowledgments 

We thank Frank van den Bosch, Andrey Kravtsov, Ariyeh Maller, Erin Sheldon, 
Risa Wechsler, David Weinberg, and Andrew Zentner for useful discussions and comments.

A.~A.~B. acknowledges support by NASA grant NAG5-11669 and NSF grant PHY-0101738.  
A.~A.~B also acknowledges the hospitality of the Aspen Center for Physics, where some 
of this work was completed.  
M.~R.~B. was partly supported during the time of this research by NSF grant AST-0607701,
NSF grant AST-0428465, and GALEX Archive Research Grant \#38.
R.~S. and S.~P. acknowledge support by NSF grant AST-0607747.
D.~W.~H. acknowledges support by NSF grant AST-0428465.

Funding for the SDSS and SDSS-II has been provided by the Alfred P. Sloan Foundation, the Participating Institutions, the National Science Foundation, the U.S. Department of Energy, the National Aeronautics and Space Administration, the Japanese Monbukagakusho, the Max Planck Society, and the Higher Education Funding Council for England. The SDSS Web Site is http://www.sdss.org/.

The SDSS is managed by the Astrophysical Research Consortium for the Participating Institutions. The Participating Institutions are the American Museum of Natural History, Astrophysical Institute Potsdam, University of Basel, Cambridge University, Case Western Reserve University, University of Chicago, Drexel University, Fermilab, the Institute for Advanced Study, the Japan Participation Group, Johns Hopkins University, the Joint Institute for Nuclear Astrophysics, the Kavli Institute for Particle Astrophysics and Cosmology, the Korean Scientist Group, the Chinese Academy of Sciences (LAMOST), Los Alamos National Laboratory, the Max-Planck-Institute for Astronomy (MPA), the Max-Planck-Institute for Astrophysics (MPIA), New Mexico State University, The Ohio State University, University of Pittsburgh, University of Portsmouth, Princeton University, the United States Naval Observatory, and the University of Washington.



\def\baselinestretch{1}

\bibliographystyle{/home/users/aberlind/TEX/apj}
\bibliography{/home/users/aberlind/TEX/bibtex.master}


\end{document}